\documentclass{article}
\usepackage{spconf,amsmath,graphicx}
\usepackage[OT1]{fontenc}
\usepackage{bm}
\usepackage{amssymb}
\usepackage{amsmath}
\usepackage{booktabs}
\usepackage{multirow}
\usepackage[colorlinks,
            linkcolor=black,
            anchorcolor=black,
            citecolor=black,
            urlcolor=black
            ]{hyperref}

\title{A NOVEL CROSS-LINGUAL VOICE CLONING approach with \\ A few text-free samples}

\name{Xinyong Zhou$^{1}$, Hao Che$^{2}$, Xiaorui Wang$^{2}$, Lei Xie$^{1}$}
\address{$^{1}$Shaanxi Provincial Key Lab of Speech and Image Information Processing, \\
School of Computer Science, Northwestern Polytechnical University, Xi’an, China\\
$^{2}$Kwai, Beijing, China\\}

\begin{document}
%
\maketitle
\begin{abstract}
This paper proposes a novel cross-lingual voice cloning framework by utilizing
bottleneck (BN) features obtained from speaker-independent automatic speech recognition system
in the target language.
Firstly, we use audio-text pairs of a single speaker in the target language
to train a latent prosody model, 
which models the relationships between the text and the BN features.
Then the acoustic model, which translates the BN features to the acoustic features,
is trained with multi-speaker's audio data in the target language. 
Finally, the acoustic model is fine-tuned with target speaker's speech
in the original language without the corresponding texts,
as the BN features are served as the bridge.
Our approach has the following advantages: 1) no recordings from bilingual speakers are required;
2) audio-text pairs are not required for acoustic model training;
3) no extra complicated modules are needed to encode speaker or language.
Experimental results show that, with only several minutes of audio from a new English speaker, our
proposed system can synthesize this speaker's Mandarin speech with decent naturalness and speaker similarity.

\end{abstract}


\section{Introduction}
\label{sec:intro}

In recent years, with the rapid development of speech synthesis, customized or personalized service 
such as voice cloning has drawn much interest.
Voice cloning aims to learn the voice of an unseen speaker from a few samples.
With a more ambitious goal, \textit{cross-lingual} voice cloning is to learn the voice from speech in other languages
and to synthesize speech in a specific language not spoken by the target speaker. 
The technology can benefit various fields such as speech translation and personalized 
computer-aided language learning.

There are some prior works focus on cross-lingual text-to-speech (TTS). 
Xie et al. \cite{xie2016kl} propose a Kullback-Leibler divergence and deep neural network (DNN) based approach 
to cross-lingual TTS training.
Li et al. \cite{li2016multi} proposed a multilingual parametric neural TTS system, which used a unified input 
representation and shared parameters across languages.
In \cite{ming2017light}, Ming et al. described a bilingual Chinese and English neural TTS system trained on speech 
from a bilingual speaker, aiming to synthesize speech of both languages using same voice.
Nachmani et al. \cite{nachmani2019unsupervised} presented a multilingual neural TTS model which supports voice
cloning across English, Spanish, and German. It used language-specific text and speaker encoders.
Xue et al. \cite{xue2019building} built a mixed-lingual TTS system with only monolingual data by adding speaker
embedding and phoneme embedding.
In \cite{sun2016personalized}, Sun et al. presented a cross-lingual TTS system using Phonetic Posteriorgrams (PPGs)
with decent voice cloning performance.
However, in this approach, a large amount of target speaker's speech is required to train a voice conversion model.
The above existing voice cloning approaches face some obvious drawbacks in real applications:
1) such as the need of recordings from bilingual speakers, or a large amount of multi-speaker audio-text pairs;
2) they need to design a specific method of phoneme sharing cross different languages;
3) they add extra modules to encode speaker and language, which complicates 
the building pipeline and may be hard to train.

Inspired by \cite{sun2016personalized}, we propose a novel cross-lingual voice cloning framework using
BN features as a bridge across speakers and language boundaries.
Suppose we have a on-the-shelf DNN-based speech recognition engine for the target language, 
and the BN features are the representation from the last hidden layer before the softmax.
Previous studies have consistently shown that BN features 
are less language-dependent \cite{geng2016multilingual} and are more smoothness on the pronunciation space 
than PPGs, since PPGs are directly defined on the phone set of the training language.


\begin{figure*}[htb]
  \centering
  \includegraphics[scale=0.84]{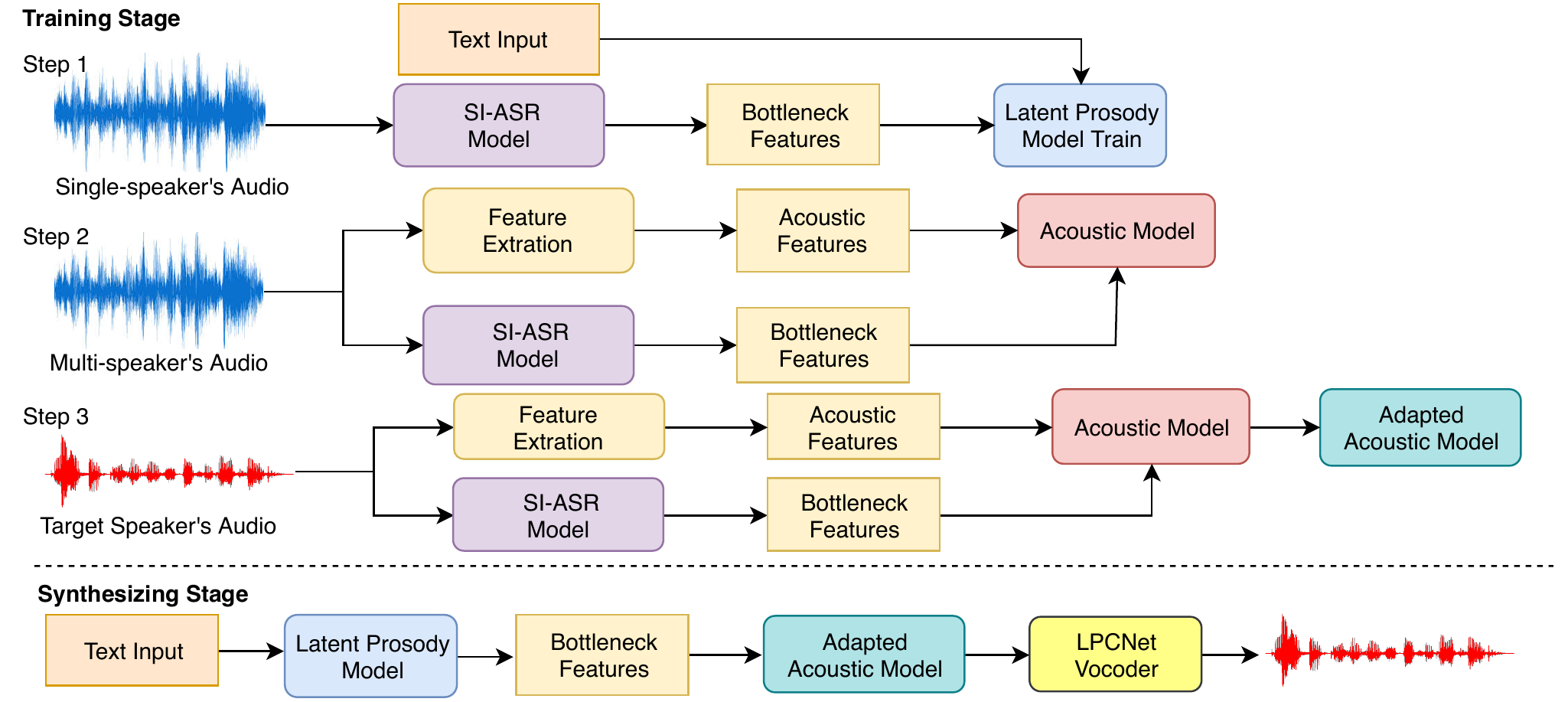}
  \caption{Architecture of framework.}
  \label{fig:our}
\end{figure*}

The proposed voicing cloning framework consists of two parts: latent prosody 
model and acoustic model. Firstly, audio-text pairs from a single speaker in the target 
language is used to train a Tacotron2-based \cite{shen2018natural} latent prosody model, which takes text sequence 
as input, predicting the corresponding BN features with automatic time alignment. 
Secondly, the CBHG-structured \cite{wang2017tacotron} acoustic model is trained with multi-speaker’s audio 
data in the target language, which translates BN features to acoustic features. 
For an unseen speaker, the acoustic model is fine-tuned using a few audio samples of this 
speaker without the need of the corresponding texts, as the input BN features to the 
acoustic model are from the output of the BN extractor. 
In the synthesis stage, for any text in the target language, 
the corresponding BN features can be predicted by the latent prosody model, 
and then the acoustic model predicts the acoustic features. 
Finally, given the acoustic features, a speaker-independent neural vocoder 
is used to synthesize the speech that sounds like the target speaker’s voice.
The advantages of our approach are as follows. 1) No recordings from bilingual speakers are required, 
where such data are hard to collect. 
2) Audio-text pairs are not required to train and fine-tune the acoustic model, 
as our approach is text-free. 
3) Our approach is simple and trainable as no extra complicated modules are needed to encode speaker or language.

The paper is organized as follows: 
Section 2 describes our approach for cross-lingual voice cloning in detail. 
Section 3 introduces the experiments and Section 4 presents subjective evaluation 
results. In Section 5, we give a brief summary and mention our future work.

\section{system architecture}
\label{sec:method}

\subsection{Bottleneck Features}
\label{ssec:stage1}

As a narrow hidden layer right before the softmax layer in DNN, the bottleneck layer 
creates a constriction in the network that forces the information pertinent to classification 
into a compact feature representation.
The work in \cite{geng2016multilingual} has shown that the BN features extracted from ASR trained 
using monolingual data perform quite well on language identification task (classifying multiple languages). 
This indicates that BN features are language-independent to some extent and
have the potential of cross-lingual task.
What's more, the BN extractor used in this paper is trained with a large-scale ASR corpus 
(containing tens of thousands of speakers), thus can be considered as speaker-independent.

\subsection{Latent Prosody Model}
\label{ssec:text-to-bn}

The latent prosody model predicts the corresponding BN features with a sequence of phoneme as input,
shown as the training step 1 in Figure \ref{fig:our}.
The model is based on Tacotron2 composed of encoder, 
attention and decoder.
The encoder takes the text input sequence \bm{$x$} of
length $L$ as an input, and learns a continuous sequential representation \bm{$h$}.
\begin{equation}
  \label{equ:encoder}
  \bm{h} = encoder(\bm{x}).
\end{equation}
The location-sensitive attention \cite{chorowski2015attention} is used as attention module, 
which uses cumulative attention weights from previous decoder time steps as an additional feature.
The decoder is an autoregressive recurrent neural network which consists of 2 uni-directional
LSTM layers. At each output time step $t$, attention and decoder word together in the following
manner: 
\begin{align}
  \label{equ:attention}
  \alpha_t &= attention(s_{t-1}, \alpha_{t-1},\bm{h}), \\
  c_t      &= \sum_{j=1}^{L} \alpha_{t,j}h_j,          \\
  y_t      &= decoder(s_{t-1}, c_t).                   
\end{align} 
where $s_{t-1}$ is the $(t-1)$-th state of the decoder, $\alpha_t \in \mathbb{R}^L$ are the 
attention weights and $c_t$ is the context vector. The decoder takes the previous hidden state
$s_{t-1}$ and the current context vector $c_t$ as inputs and generates the current output $y_t$.
We minimize the ground truth BN features and the predicted BN features with L2 loss.

\subsection{Acoustic Model}
\label{ssec:avm}

The acoustic model is based on CBHG,
consisting of a bank of 1-D convolutional filters, followed by highway networks and a bidirectional
GRU recurrent neural network.
Besides, two pre-net layers is added to improve model's generalization ability.
For a given utterance from the corpus, $t$ denotes the 
frame index of this sequence. The input is the BN features ($B_1$,\dots,$B_t$,\dots,$B_N$). 
The target value of the output layer is the acoustic features sequence ($Y_1^T$,\dots,$Y_t^T$,\dots,$Y_N^T$). 
The predicted value of the output layer is ($Y_1^R$,\dots,$Y_t^R$,\dots,$Y_N^R$). 
The cost function of training process is defined as follows:
\begin{equation}
  \label{equ:loss}
  min\sum_{t=1}^{N}\left \| Y_t^R-Y_t^T\right \|.
\end{equation}

\subsection{Speaker Adaptation}
\label{ssec:stage3}

As shown in the training step 3 in Figure \ref{fig:our}, 
given a few speech samples of target speaker, we extract the BN features and the acoustic features firstly.
The the acoustic model is fine-tuned using these features without the need of the corresponding texts.
For a given utterance, the corresponding BN features are predicted by the latent prosody model, 
then the BN features are fed into the adapted acoustic model to predict acoustic features. 
Finally, a speaker-independent neural vocoder is used to synthesize the speech.


\section{EXPERIMENTS}
\label{sec:experiment} 
In our work, Mandarin and English are chosen as target language and non-target language respectively.
We aim to synthesize one speaker's speech in Mandarin (with arbitrary textual input),
given just a few English speech samples of this speaker, without the corresponding texts.
As illustrated in Figure \ref{fig:our},
the proposed framework is divided into training stage and synthesizing stage.
In training stage, the latent prosody model
and the acoustic model can be trained in parallel. 
Then the acoustic model is fine-tuned using a few audio samples of target speaker.
In the synthesis stage, for any text in the target language, 
the corresponding BN features 
can be predicted by the latent prosody model, and then the acoustic model predicts the 
acoustic features. 
Finally, a speaker-independent LPCNet vocoder \cite{valin2019lpcnet} is used to synthesize speech.

\subsection{Dataset}
\label{ssec:data}

The training data of the latent prosody model is DB-1 corpus from Databaker technology, 
which contains 12 hours of Mandarin female data.
For the acoustic model training, THCHS-30 (an open Chinese speech database published by 
Center for Speech and  Language Technology (CSLT) at Tsinghua University) \cite{THCHS30_2015} is used, which
contains 60 speakers, each of whom has about 250 sentences.
The signals are sampled at 16kHz with mono channel, windowed with 25 ms and shifted every 10 ms. 
Acoustic feature includes 30-dimensional bark-frequency cepstral coefficients (BFCC), 2 pitch parameters 
(period, correlation).
The dimension of BN features is 512.

\subsection{Baseline approach}
\label{ssec:baseline}

The baseline approach is a speaker adaptation training approach.
The model is based on Tacotron2 architecture, which predicts the acoustic features
from the phoneme sequence directly.
Firstly, the model is trained using THCHS-30 data, then fine-tuned with a few target speaker's audio-text pairs.
The vocoder used is same as our approach.
Since our training data is only Mandarin, this approach can only be applied to Mandarin speaker adaptation.
We compared the effect of our proposed and baseline approaches on Mandarin speakers.

\subsection{Experimental Setup}
\label{ssec:exp_setup}

The ASR model is based on time-delay neural network-long short term memory (TDNN-LSTM) \cite{peddinti2017low}.
We take the output of the last LSTM layer (near to softmax layer) as BN features.
The SI-ASR system is implemented using Kaldi speech recognition toolkit \cite{povey2011kaldi}.
For latent prosody model and baseline training, we follow the specifications mentioned in Tacotron2 \cite{shen2018natural}.
We use the Adam optimizer \cite{kingma2014adam} with learning rate decay, which
starts from 1e-3 and is reduced to 1e-5 after 50k steps.
The network is trained with a batch size of 16 with an NVIDIA GTX1080Ti GPU. 
Transcripts are converted to the corresponding phoneme sequences using grapheme-to-phoneme (G2P) library. 
For Chinese text, we first perform word segmentation which separates words and phrases with specific
symbols in order to improve speech fluency. The phoneme sequences are fed to the encoder of latent prosody
model as input.
Acoustic model is pre-trained for 200k steps with L1 loss and a batch size of 32,
then fine-tuned for 4k steps with target speaker's data.
The specific hyperparameters are consistent with \cite{wang2017tacotron}.


\subsection{Evaluation}
\label{ssec:subjective}
One Mandarin male (MM) speaker and one Mandarin female (MF) speaker are chosen as target speakers to compared
the effects of our proposed and baseline approaches.
\begin{table}[htb]
  \centering
  \begin{tabular}{cccc}
  \hline
  \multirow{2}{*}{\begin{tabular}[c]{@{}c@{}}Target\\ Speaker\end{tabular}} & \multicolumn{3}{c}{Number of sentences} \\ \cline{2-4} 
   & 50 & 100 & 200 \\ \hline
  MF (baseline) & 3.15$\pm$0.12 & 3.26$\pm$0.11 & 3.56$\pm$0.09 \\
  MF (our) & 3.82$\pm$0.11 & 3.95$\pm$0.10 & 3.98$\pm$0.08 \\ \hline
  MM (baseline) &  2.85$\pm$0.13 & 3.04$\pm$0.09 & 3.11$\pm$0.10 \\
  MM (our) &  3.68$\pm$0.09 & 3.73$\pm$0.12 & 3.75$\pm$0.09 \\ \hline
  EF (our) & 3.72$\pm$0.08 & 3.82$\pm$0.11 & 3.88$\pm$0.09 \\
  EM (our) & 3.63$\pm$0.06 & 3.71$\pm$0.07 & 3.72$\pm$0.08 \\\hline
  \end{tabular}
  \caption{Speech naturalness Mean Opinion Score (MOS) with 95\% confidence intervals.}
  \label{mos_naturalness}
\end{table}
In addition, one English male (EM) speaker and one English female (EF) speaker are used to evaluate 
the proposed approach on cross-lingual voice cloning.
For measuring the naturalness and speaker similarity of synthesize speech, we conducted a 
Mean Opinion Score (MOS) test as subjective evaluations.
For each set of experiments, 50, 100 and 200 sentences of target speaker are used to fine-tune acoustic model respectively and 
randomly select 30 sentences (not in the training set) for testing.
\renewcommand{\thefootnote}{\fnsymbol{footnote}}
\footnote{\href{https://xinge333.github.io/speaker\_adaptation\_demo}
{Audio samples: https://xinge333.github.io/speaker\_adaptation\_demo}}



\subsubsection{Speech naturalness}
\label{ssec:naturalness}
We invite 20 listeners to participate in the subjective tests,
whose first language is Chinese and are well educated in English.
The subjects are asked to rate the naturalness of generated utterances on a five-point Likert scale 
(1:Bad, 2:Poor, 3:Fair, 4:Good, 5:Excellent). 
The results are shown in Table \ref{mos_naturalness}, which indicates our proposed approach performs better than
baseline in naturalness on Mandarin.
In the case of extremely limited data, the speech synthesized by our approach is still clear and stable,
but baseline is not.
What's more, our approach achieve satisfactory performance in speech quality and naturalness
for English speakers.


\subsubsection{Speech similarity}
\label{ssec:similarity}
In similarity test, a subject is presented with a pair of utterances
comprises a real utterance recorded by a speaker and another synthesized utterance from the same speaker. 
The similarity MOS test uses five-scale-score for evaluation (1: Not at all
similar, 2: Slightly similar, 3: Moderately similar, 4: Very similar, 5: Extremely similar). 
As shown in Table \ref{mos_similarity}, 
our proposed approach performs better than baseline in speaker similarity on Mandarin.
The similarity MOS of English speakers are both above 3.5, which demonstrates the model 
can primely generalize to the new cross-lingual speakers.




\begin{table}[htb]
  \centering
  \begin{tabular}{cccc}
  \hline
  \multirow{2}{*}{\begin{tabular}[c]{@{}c@{}}Target\\ Speaker\end{tabular}} & \multicolumn{3}{c}{Number of sentences} \\ \cline{2-4} 
   & 50 & 100 & 200 \\ \hline
   MF (baseline) & 3.85$\pm$0.05 & 3.92$\pm$0.10 & 4.06$\pm$0.08 \\
   MF (our) & 3.97$\pm$0.08 & 4.12$\pm$0.11 & 4.15$\pm$0.09 \\\hline
   MM (baseline) &  3.25$\pm$0.12 & 3.34$\pm$0.13 & 3.41$\pm$0.11 \\
   MM (our) &  3.74$\pm$0.11 & 3.86$\pm$0.06 & 3.89$\pm$0.10 \\\hline
   EF (our) & 4.02$\pm$0.09 & 4.10$\pm$0.08 & 4.11$\pm$0.07 \\
   EM (our) & 3.75$\pm$0.12 & 3.81$\pm$0.09 & 3.87$\pm$0.07 \\\hline
  \end{tabular}
  \caption{Speech similarity Mean Opinion Score (MOS) with 95\% confidence intervals.}
  \label{mos_similarity}
\end{table}
\subsection{Analysis}
\label{ssec:Analysis}
We analyze why our proposed approach can realize cross-lingual voice cloning using
limited audio samples.
We think that the most important thing is the utilization of BN features in our approach.
Firstly, the BN features are language-independent and speaker-independent,
which makes the pronunciation space of different languages and speakers can be represented uniformly.
Secondly, compared with other acoustic feature such as mel-frequency spectrogram, 
BN feature as a high-level feature is insensitive to background noise, speaker channel, accent and gender.
Finally, since BN feature is close to the softmax layer of ASR model, we think it is acoustics
weakly dependent and mainly contains linguistic feature and prosody information.
Therefore, it is easy for the latent prosody model to learn the mapping between text and BN feature.
When the acoustic model is fine-tuned, the model does not need to consider prosody information of target speaker.
But in baseline approach,
the model needs to learn not only the acoustics-related information of the target speaker but also 
the speaking style information including prosody.
Thus even with extremely limited data, our approach can still acquire a stable pronunciation effect
while learning the target speaker's voice.




\section{CONCLUSIONS}
\label{sec:conclusion}

In this paper, we present a cross-lingual voice cloning approach.
BN features obtained by SI-ASR model are used as
a bridge across speakers and language boundaries.
The relationships between text and BN features are modeled by the latent prosody model. 
The acoustic model learns the translation from BN features to 
acoustic features. 
The acoustic model is fine-tuned with a few samples of the target speaker to realize voice cloning.
This system can generate speech of arbitrary utterance of target language in cross-lingual speakers' voice. 
We verify that with small amount of audio data, our proposed approach can well handle cross-lingual tasks.
And in intra-lingual tasks,
our proposed approach also performs better than baseline approach in naturalness and similarity.

Because BN features lost energy information,
the synthesized speech is lack of expressiveness in stress and mood words.
Besides, the speaking style of synthetic speech is fixed, as the latent prosody model is trained
using a single speaker's corpus.
In the future, more research will be do on synthesizing expressive personalized speech and
transferring the prosodic style of the target speaker while performing voice cloning.
\bibliographystyle{IEEEbib}
\bibliography{strings,refs}

\end{document}